\documentclass[aps,prl,showpacs,twocolumn,groupedaddress]{revtex4}

\usepackage{graphicx,bbm,amsthm, amsmath, amssymb}
\usepackage{color}

\def\tr{\,{\rm tr}\,}
\def\ket#1{|#1\rangle}

\def\braket#1#2{\langle #1 | #2 \rangle}
\def\rx#1#2{R_{\rm x}^{(#1)}(#2)}
\def\ry#1#2{R_{\rm y}^{(#1)}(#2)}

\def\tit#1{}
\def\etal#1{ {\em et.al.}}

\begin{document}

\title{How many CNOT gates does it take to generate a three-qubit state ?}

\author{Marko \v Znidari\v c$^1$, Olivier Giraud$^2$ and Bertrand Georgeot$^2$}

\affiliation{
${}^1$ Department of Physics, Faculty of Mathematics and Physics, 
University of Ljubljana, 
SI-1000 Ljubljana, Slovenia\\
${}^2$ Laboratoire de Physique Th\'eorique, 
Universit\'e de Toulouse, CNRS, 31062 Toulouse, France}

\date{November 26, 2007}

\begin{abstract}
The number of two-qubit gates required to transform deterministically 
a three-qubit pure
quantum state
into another is discussed. We show that any state can be prepared from 
a product state using at most three CNOT gates, and that, starting from the GHZ state, only two suffice. As a consequence, any three-qubit state can be transformed into any other using at most four CNOT gates.  Generalizations to other
two-qubit gates are also discussed.
\end{abstract}


\maketitle

Quantum information and computation (see e.g.~\cite{nielsen}) is usually
described using qubits as elementary units of information which are
manipulated through quantum operators. In most practical implementations,
such operators have to be realized as sequences of
local transformations acting on a few qubits at a time.  Whereas one-qubit
gates alone cannot create entanglement, it has been shown
that together with two-qubit gates they can form
universal sets, from which the set of all unitary transformations
of any number of qubits can be generated \cite{BarBenCle95}.
The complexity of a quantum algorithm is usually measured
by assessing the number of elementary gates needed to perform
the computation.
The Controlled-Not (CNOT) gate, a two-qubit gate
whose action can be written
$|00\rangle \rightarrow |00\rangle$, $|01\rangle \rightarrow |01\rangle$,
$|10\rangle \rightarrow |11\rangle$, $|11\rangle \rightarrow |10\rangle$,
is one of the most widely used both for theory and implementations.
It can be shown that the CNOT gate together
with one-qubit gates is a universal set \cite{BarBenCle95}.
Experimental implementations of a CNOT gate (or the equivalent
controlled phase-flip) have been recently reported
using e.g. atom-photon interaction in cavities \cite{haroche}, 
linear optics \cite{linopt}, superconducting qubits \cite{nakamura,delft} or
ion traps \cite{wineland,blatt}.  While large size quantum computers are still 
far away, small platforms of a few qubits exist or can be envisioned in the
framework of these existing experimental techniques. In most such 
implementations, two-qubit gates such as the CNOT are much more demanding that 
one-qubit gates.  

Theoretical quantum computation has been usually focused on assessing
the number of elementary gates to build a given unitary operator performing 
a given computation.  Some works have tried to focus on two-qubit
gates and to minimize their number in order to build a given
unitary transformation
for several qubits \cite{U3cnot,unitary}.  Still, unitary transformations in
many applications are a tool to transform an initial state to a given state.
It seems therefore natural to try and assess how costly this process is in
itself. 
In this paper, we thus study the minimal number
of two-qubit gates needed to change a given quantum state to obtain
another one. Of course, this number is necessarily 
upper bounded by the number required for a general unitary transformation.
We focus on the case of two and three qubits.  For two
qubits, we generalize the result of \cite{2qubit00} and show that one 
CNOT is enough to go from any given pure state to any other one.  For
three qubits, we show that three CNOTs are enough to go from $|000\rangle$
to any other pure state, and that two CNOTs suffice 
if one starts from the GHZ state
$(|000\rangle +|111\rangle)/\sqrt{2} $.  A corollary of the latter is
thus that four CNOTs are enough
to go from any pure state to any other pure state. 
The number of CNOT gates required to go from a state to another defines 
a discrete distance
on the Hilbert space. Given any fixed state $\ket{\psi}$, the Hilbert space
 can be partitioned according to the distance to $\ket{\psi}$.
It is known that if stochastic one-qubit 
operations are used,
entanglement of three \cite{SLOCC3} and four 
\cite{SLOCC4} qubits fall into respectively two and nine
different classes. Our classification according to the number of CNOTs
is different, although there are some relations.  Our results 
generalize to other universal two-qubit gates, in particular to the
iSWAP gate which has been shown to be implementable for
superconducting qubits \cite{MajChoGam07}.

We consider pure states belonging to the $2^n$-dimensional Hilbert space 
$\mathbb{C}^{2^n}$.
The space of normalized quantum states is the sphere $S^{2^{n+1}-1}$.
As the cost of one-qubit gates is negligible, we are interested only in
equivalence classes of states modulo local unitary transformations
(LU). We thus consider
the sets $\mathcal{E}_n=S^{2^{n+1}-1}/U(2)^{n}$ of states nonequivalent 
under LU. In the case of two and three qubits the dimension of
$\mathcal{E}_2$ and  $\mathcal{E}_3$ was determined 
in \cite{Schlienz96,Linden98} and their topology 
has been described in \cite{WalGlaLoc05}.
Throughout the paper we will make use of the one-qubit LU operations
$R_{\rm j}^{(k)}(\xi)=\exp{( - {\rm i} \xi \sigma_{\rm j}^{(k)})}$ 
where the $\sigma_{\rm j}^{(k)}$ are the Pauli matrices acting on
qubit $k$. In particular, the operation $R_{\rm y}^{(k)}(\xi)$ 
corresponds to a rotation of the qubit
$\cos(\varphi)\ket{0}+\sin(\varphi)\ket{1}\mapsto 
\cos(\varphi + \xi)\ket{0}+\sin(\varphi+\xi)\ket{1}$.

{\bf Two-qubit states.}
Let us first consider the two-qubit case. It has
been shown in~\cite{2qubit00} that one can transform an arbitrary two-qubit 
state $\ket{\psi}$ to $\ket{00}$ by using only one CNOT. Here we 
prove that the same holds for two general two-qubit states $\ket{\psi}$ and 
$\ket{\psi'}$. 

{\em Proof:} Since $\mathcal{E}_2$ is homeomorphic 
to $[0,1]$, only one parameter (e.g., one Schmidt coefficient)
characterizes a state up to LU. More precisely,
by LU each two-qubit state can be brought to the canonical form 
$\ket{\psi}=\cos{\varphi}\ket{00}+\sin{\varphi}\ket{11}$, which is just Schmidt decomposition. We want to transform state $\ket{\psi}$ with 
parameter $\varphi$ to state $\ket{\psi'}$ with parameter $\varphi'$. 
When $\varphi'\neq\varphi$, we need at least one CNOT. It turns out that one CNOT is in fact sufficient, as 
can be easily seen by checking that the relation 
$\ry{1}{-\varphi}\rm{CNOT}_{12}\ry{1}{\varphi'}\ket{\psi}=\ket{\psi'}$ holds 
(by convention, qubit 1 is the leftmost one).
\qed

In contrast, one needs three CNOTs in general to construct a specific
two-qubit unitary transformation~\cite{U3cnot}. Transforming
one state to another is thus clearly easier.

We now turn to the three-qubit case.

{\bf Classification with respect to $\ket{000}$.}
We start with the case where we want to prepare 
a state $\ket{\psi}$ from $\ket{000}$.
 The distance (in number of CNOTs)  from 
$\ket{000}$ to $\ket{\psi}$ is a criterion for the difficulty to prepare
$\ket{\psi}$. We will show that this
distance partitions the 
Hilbert space into four classes, and that any state can be prepared from
$\ket{000}$ using three CNOT gates or less. We will examine each of these 
four classes in turn.

{\bf Class 0:} One needs zero CNOT gates to transform $\ket{\psi}$ 
to $\ket{000}$ iff the state is of the product form 
$\ket{\psi}=\ket{\alpha \beta \gamma}$,
where $\ket{\alpha},\ket{\beta},\ket{\gamma}$ are 
normalized single qubit states
(this is trivial, since
only LU are used).

{\bf Class 1:} One needs one CNOT iff the state is of the form 
$\ket{\psi}=\ket{\alpha}_1\ket{\chi}_{23}$ (i.e., it is bi-separable), 
where $\ket{\chi}_{23}$ is an arbitrary entangled state of the last two
qubits \cite{relabeling}.

{\em Proof:} By LU $\ket{\psi}$ can be transformed into canonical form 
$\ket{\psi}\stackrel{{\rm \tiny LU}}{=} \ket{0}(\cos{\varphi}\ket{00}+\sin{\varphi}\ket{11})$. 
Applying CNOT $_{23}$ to this canonical form
we obtain $\ket{0}(\cos{\varphi}\ket{00}+\sin{\varphi}\ket{10})$, 
i.e., state $\ket{0}(\cos{\varphi}\ket{0}+\sin{\varphi}\ket{1})\ket{0}$ which 
is in class 0. We can therefore reach $\ket{000}$ in a single CNOT step. 
Conversely,
applying one CNOT gate on a state from class 0 
the state reached is bi-separable and therefore all states 
that need 1 step to get from $\ket{000}$ are of the above form.\qed

In their canonical form states from class 1 can be parametrized by a 
single real parameter $\varphi$.

{\bf Class 2:} One needs two CNOT gates iff the state is of the form 
$\ket{\psi}=\cos{\varphi}\ket{\alpha \beta \gamma}
+\sin{\varphi}\ket{\alpha_\perp \beta' \gamma'}$, 
with $\braket{\alpha}{\alpha_\perp}=0$, $|\braket{\beta}{\beta'}|<1$ and 
$|\braket{\gamma}{\gamma'}|<1$ (if $|\braket{\beta}{\beta'}|$ or 
$|\braket{\gamma}{\gamma'}|$ are equal to 1 then $\ket{\psi}$ 
belongs to class 1).

{\em Proof:} We first bring the state 
by LU to the canonical form $\ket{\psi} \stackrel{{\rm \tiny LU}}{=}
\cos{\varphi} \ket{000}+\sin{\varphi}\ket{1\beta\gamma}$. 
If the phases are absorbed into the definition of local bases, 
$\ket{\gamma}$ can be written as $\cos{\xi}\ket{0}+\sin{\xi}\ket{1}$.
The rotation $R_{\rm y}^{(3)}\left(\frac{\pi}{4}-\frac{\xi}{2}\right)$
followed by a CNOT$_{13}$ yields a state
$(\cos{\varphi}\ket{00}+\sin{\varphi}\ket{1\beta})\ket{\gamma'}$, 
which is a state of class 1 from which we can reach state $\ket{000}$ 
in a single step.

To prove the converse, we have to show that by using 
two CNOT gates one can reach only states of the form 
$\ket{\psi}=\cos{\varphi}\ket{\alpha \beta \gamma}+\sin{\varphi}
\ket{\alpha_\perp \beta' \gamma'}$, or states in class 0 or class 1. 
Starting from class 0, we are 
in class 1 after one step.  Any class 1 state can be written as
$\ket{\alpha}(\cos{\varphi} \ket{0\gamma}+{\rm e}^{{\rm i}\xi} \sin{\varphi}\ket{1\gamma'})$. 
Applying CNOT$_{23}$ or CNOT$_{32}$ we get a state in class 0 or class 1. Applying CNOT$_{21}$ 
on the other hand we get $\cos{\varphi}\ket{\alpha 0\gamma}
+{\rm e}^{{\rm i}\xi}\sin{\varphi}\ket{\overline{\alpha}\,1\gamma'}$, 
where $\ket{\overline{\alpha}}=\sigma_{\rm x}\ket{\alpha}$,
which is indeed of the canonical form of class 2 states.
Last possibility is applying CNOT$_{12}$. In this case it is better to write 
our state in the 
basis $\ket{\pm}=(\ket{0}\pm \ket{1})/\sqrt{2}$ for the second qubit. 
That is, any class 1 state can be written as 
$\ket{\alpha}(\cos{\varphi} \ket{-\gamma}+{\rm e}^{{\rm i}\xi} \sin{\varphi}\ket{+\gamma'})$. Writing 
$\ket{\alpha}=\cos{\varphi'}\ket{0}+{\rm e}^{{\rm i} \xi'}\sin{\varphi'}\ket{1}$, 
we then get after applying CNOT$_{12}$ state 
$\cos{\varphi}(\cos{\varphi'}\ket{0}-{\rm e}^{{\rm i}\xi'}\sin{\varphi'}\ket{1})
\ket{- \gamma}+\sin{\varphi} {\rm e}^{{\rm i}\xi} (\cos{\varphi'}
\ket{0}+{\rm e}^{{\rm i}\xi'}\sin{\varphi'}\ket{1})\ket{+ \gamma'}$, 
which is again of the canonical form expected. For CNOT$_{13}$ or CNOT$_{31}$ 
the argument is similar.\qed

One needs 3 real parameters to describe states of class 2 in their canonical 
form. Note that class 2 states constitute a subset of GHZ type states which are of (unnormalized) form $\ket{\alpha \beta \gamma}+\ket{\alpha' \beta' \gamma'}$
\cite{SLOCC3}.

{\bf Class 3:} One needs three CNOT gates iff
a state is not in class 0, 1 or 2.

{\em Proof:}
States not in the previous classes
are of two types: (i) W-like states 
(according to the classification in \cite{SLOCC3}) 
for which the range of the reduced density matrix of qubits 2 and 3 contains only one product state.
Such states are of the (unnormalized) form 
$\ket{\psi}=\ket{\alpha \beta \gamma}+\ket{\alpha'}\ket{\chi}_{23}$, where $\ket{\chi}_{23}$ is entangled and orthogonal to 
$\ket{\beta \gamma}$. Under LU they can be written in the following canonical form \cite{SLOCC3}
\begin{equation}
\ket{\psi} \stackrel{{\rm LU}}{=} \cos{\varphi}\ket{000}+\sin{\varphi}
\ket{\alpha} (\cos{\varphi'} \ket{10}+\sin{\varphi'}\ket{01}),
\label{eq:3W}
\end{equation}
with $\ket{\alpha}=\cos{\xi}\ket{0}+\sin{\xi}\ket{1}$ and, 
(ii) GHZ-like states with the canonical form
\begin{equation}
\ket{\psi} \stackrel{{\rm LU}}{=} a \ket{000}+{\rm e}^{{\rm i}\xi} b \ket{\alpha \beta \gamma},
\label{eq:3GHZ}
\end{equation}
where $\ket{\alpha}$, $\ket{\beta}$ and $\ket{\gamma}$ are real 
single qubit states parametrized by one parameter each and $a,b$ 
are real parameters, one of which is fixed by normalization
\cite{markoremark}.  To 
exclude class 2 states we must demand that none of $\ket{\alpha}$, 
$\ket{\beta}$ and $\ket{\gamma}$ be equal to $\ket{1}$. To exclude class 1 and 0 states in 
(\ref{eq:3W}),(\ref{eq:3GHZ}) $\rho_{23}$ must be of rank 2. W-like 
states (\ref{eq:3W}) require 3 parameters.
GHZ-like (\ref{eq:3GHZ}) states need 5, and thus are the generic states.

First we show that by using single CNOT 
one can transform class 3 states to class 2. For W-like states (\ref{eq:3W})
we just have to apply CNOT$_{23}$ to the canonical 
form (\ref{eq:3W}) and we immediately get 
$\cos{\varphi}\ket{000}+\sin{\varphi}\ket{\alpha}(\cos{\varphi'}\ket{1}
+\sin{\varphi'}\ket{0})\ket{1}$ which is of class 2 (states 
on the third qubit are orthogonal). For GHZ-like states \eqref{eq:3GHZ} 
it is a bit more work. Note that GHZ-type states can be, by rearranging
 terms and after LU, written as $\cos{\varphi}\ket{000}
+\sin{\varphi}\ket{1}\ket{\chi}_{\rm 23}$ (expanding $\ket{0}$ on the first qubit 
in \eqref{eq:3GHZ} into $\ket{\alpha}$ and
 $\ket{\alpha_\perp}$ and adding $\ket{\alpha}$ part to the second term), where $\ket{\chi}_{\rm 23}$ 
can in turn be expanded as 
$\ket{\chi}_{\rm 23}=\cos{\varphi'}\ket{0\delta}+\sin{\varphi'}\ket{1\delta'})$ 
with $|\braket{\delta}{\delta'}|< 1$ (otherwise state would be in class 2). Finally, rotating 
third qubit brings the state to 
$\cos{\varphi} \ket{00\gamma''}+\sin{\varphi}\ket{1}(\cos{\varphi'}\ket{00}+
{\rm e}^{{\rm i}\xi'}\sin{\varphi'}\ket{1\gamma'})$ with real 
$\ket{\gamma'}=\cos{g'}\ket{0}+\sin{g'}\ket{1}$.
 After application of the rotation $R_{\rm y}^{(3)}\left(\frac{\pi}{4}-\frac{g'}{2}\right)$
followed by CNOT$_{23}$ we arrive at 
$\cos{\varphi}\ket{00\tilde{\gamma}''}+\sin{\varphi}\ket{1}(\cos{\varphi'}
\ket{0}+{\rm e}^{{\rm i}\xi'}\sin{\varphi'}\ket{1})\ket{\tilde{\gamma}'}$
which is of class 2.

Since the canonical forms of classes 0, 1, 2 and 3 span the whole 
Hilbert space and since the forms of classes 0, 1 and 2 are the only ones
that can be reached in 2 steps, it immediately follows that the states
of canonical forms  \eqref{eq:3W} and \eqref{eq:3GHZ} require exactly three 
steps.\qed 

Thus any state is at a distance less than or equal to 3
from $\ket{000}$.

{\bf Classification with respect to GHZ state.}
Let us now examine the distance to
the GHZ state
$\ket{{\rm GHZ}}=\left( \ket{000}+\ket{111} \right)/\sqrt{2}$.
It turns out that any state is at a distance less than 
or equal to 2 from GHZ. We use the fact that any state in $\mathcal{E}_3$
can be characterized by a set of six polynomial invariants \cite{Sud01, AciAndJan01}.
Following \cite{AciAndJan01}, we denote the first three invariants by
$I_i=\tr\rho_i^2$, $1\leq i \leq 3$, where $\rho_i$ is the reduced density matrix 
of the $i$th qubit. Any three-qubit state is equivalent under LU to its canonical form
\cite{AciAndJan01}
\begin{equation}
\label{decompcanonic3}
\lambda_0\ket{000}+\lambda_1{\rm e}^{{\rm i}\phi}\ket{100}+\lambda_2\ket{101}
+\lambda_3\ket{110}+\lambda_4\ket{111},
\end{equation}
where the six parameters $\lambda_0,\ldots, \lambda_4,\varphi$ label the state
in  $\mathcal{E}_3$ and $\sum\lambda_i^2=1$. 
If we take $0\leq\varphi\leq\pi$ the parameters in the
canonical form \eqref{decompcanonic3} are unique.

{\bf Class 0:} States at distance 0 from GHZ are states whose 
canonical form \eqref{decompcanonic3} is GHZ (this is trivial).

{\bf Class 1:} States at distance 1 from GHZ are states whose 
canonical form  \eqref{decompcanonic3} is (up to relabeling)
\begin{equation}
\frac{1}{\sqrt{2}}\ket{000}+\lambda_2\ket{101}
+\lambda_3\ket{110}+\lambda_4\ket{111},
\label{eq:I05}
\end{equation}
(with $\lambda_4\neq 1/\sqrt{2}$) i.e., states with $I_1=\frac{1}{2}$.

{\em Proof:} It is straightforward to check that the state (\ref{eq:I05})
has the same invariants (i.e. is the same up to LU) 
as ${\rm CNOT}_{23} \ry{2}{\theta_2}
\ry{3}{\theta_3}\ket{{\rm GHZ}}$ with $\theta_{2}=\frac{1}{2}\arcsin(\lambda_2\sqrt{2})$
and $\theta_{3}=\frac{1}{2}\arccos(\lambda_3\sqrt{2})$, and thus is at distance 1
from GHZ.
Conversely, if a state $\ket{\psi}$ is at distance 1 from GHZ, then one of its invariants
$I_1, I_2, I_3$ has to be the same as for GHZ. Since GHZ is symmetric under permutation
of the qubits, one can consider that the CNOT gate applied is ${\rm CNOT}_{23}$ 
(up to relabeling of the qubits before applying the CNOT). In this case the invariant 
$I_1=\frac{1}{2}$ is conserved. 
The reduced
density matrix of the first qubit of $\ket{\psi}$ therefore verifies $\tr\rho_1^2=\frac{1}{2}$
which in turn (since it is a $2\times2$ density matrix) implies that
 $\rho_{\rm 1}=\frac{1}{2} \mathbbm{1}$.
All states with this property can be written as
$\ket{\psi}=\frac{1}{\sqrt{2}}\ket{0} \ket{\chi}+\frac{1}{\sqrt{2}}
\ket{1}\ket{\chi_\perp}$, with $\braket{\chi}{\chi_\perp}=0$. The canonical 
form of such states is precisely (\ref{eq:I05}).\qed

{\bf Class 2:} All other states are at distance 2 from GHZ.

{\em Proof:} We show that all other states are at distance 1 from 
states of canonical form (\ref{eq:I05}). States  of canonical form (\ref{eq:I05})
are characterized by  $I_1=\frac{1}{2}$. Therefore we have to prove that
any state  $\ket{\psi}$ not in class 0 or 1 can be transformed to a state 
verifying $I_1=\frac{1}{2}$ with only one CNOT gate.
 The reduced density matrix of the first qubit is a $2\times 2$-matrix 
$\rho_1=\left(\begin{array}{cc}A&B\\B^{*}&1-A\end{array}\right)$. The first invariant
is thus given by $I_1=A^2+(1-A)^2+2|B|^2$. It is equal to $\frac{1}{2}$ iff $A=\frac{1}{2}$ and $B=0$, 
which (since $B$ is complex) yields three equations. 
One-parameter rotations of the qubits before applying CNOT 
yield free parameters. It turns out that a solution to 
the three equations always exists.
Indeed, the 
state $\ket{\psi}$ can be reduced either to the canonical form 
(\ref{eq:3W}) or to  the canonical form (\ref{eq:3GHZ}).
If $\ket{\psi}$ is in canonical form (\ref{eq:3W}), one can check that 
the state ${\rm CNOT}_{12} \ry{1}{\theta_1}\ry{2}{\theta_2}\ket{\psi}$, with
\begin{eqnarray}
\tan{2\theta_1}&=&\frac{\cot^2{\varphi}+\cos{2\xi}}{\sin{2\xi}} \\
\tan{2\theta_2}&=& \frac{\sin{(\xi+2\theta_1)}\cos{\varphi'}
\sin{2\varphi}}{\sin{(2\xi+2\theta_1)}\cos{2\varphi'}\sin^2{\varphi}
-\cos^2{\varphi}\sin{2\theta_1}}, \nonumber
\end{eqnarray}
is such that $I_1=\frac{1}{2}$. Let us now suppose that $\ket{\psi}$ is in
canonical form (\ref{eq:3GHZ}). The one-qubit states $\ket{\alpha}$,  $\ket{\beta}$, 
$\ket{\gamma}$ can be written in the form $\cos{\varphi}\ket{0}+\sin{\varphi}\ket{1}$
with parameters respectively $\varphi_1$, $\varphi_2$ and $\varphi_3$.
Normalization imposes that $a^2+b^2+2 a b\cos\xi \cos\varphi_1\cos\varphi_2\cos\varphi_3=1$. 
One can check that the state 
${\rm CNOT}_{12} \rx{1}{\chi}\ry{1}{\theta_1}\ry{2}{\theta_2}\ket{\psi}$, with
parameters of the rotations given by
\begin{widetext}
\begin{eqnarray}
\tan{2\theta_1} &=&
\frac{(b^2-b^4)\cos{2\varphi_1}-a^2\left(a^2-1-2b^2\cos^2\varphi_1\cos{2\varphi_2}+
2b^4\cos^2\varphi_2\sin^2(2\varphi_1)\sin^2\varphi_3\right)}
{b^2\sin(2\varphi_1)\left(1-a^2-b^2+2a^2\cos^2\varphi_2(1-a^2\sin^2\varphi_3
+b^2\cos(2\varphi_1)\sin^2\varphi_3)\right)}\nonumber\\
\tan{2\theta_2}&=&-\frac{b^2\sin{(2\varphi_1+2\theta_1)}
\sin{2\varphi_2}+2ab\cos\xi\sin{(\varphi_1+2\theta_1)}\sin{\varphi_2}\cos{\varphi_3}}
{a^2\sin{2\theta_1}
+b^2\sin{(2\varphi_1+2\theta_1)}\cos{2\varphi_2}
+2ab\cos\xi\sin{(\varphi_1+2\theta_1)}\cos{\varphi_2}\cos{\varphi_3}}\\
\tan{2\chi}&=&-\frac{2a b \sin\xi\sin\varphi_1\sin\varphi_2\cos\varphi_3(a^2\sin{2\theta_1}
-b^2\sin{(2\varphi_1+2\theta_1)})}
{2a\sin\varphi_1\sin\varphi_2(2ab^2\cos\varphi_1\cos\varphi_2
+b(a^2+b^2)\cos\xi\cos\varphi_3)},\nonumber
\end{eqnarray}
\end{widetext}
verifies $I_1=\frac{1}{2}$. \qed

The results above show that the number of CNOTs needed to transform a 
state to another state
is much less than the number to produce all unitary transformations.
Indeed, according to \cite{unitary}, one needs at least $14$ CNOT gates
to produce any three-qubit unitary transformation.  We also note that
the best available algorithm actually does not saturate the bound, needing
$20$ CNOTs \cite{unitary}.  Our procedure is explicit, and
selects a
specific unitary transformation leading from one state to the other
using less CNOTs than a general unitary transformation.

Let us now discuss the applicability of these results to physical systems.
In an experimental context, our results can be used to construct 
any desired state from the initial state which is easiest to produce
with a given system.  
In some experimental set-ups, two-qubit gates are built from a
nearest neighbor interaction (for example in \cite{nakamura, delft}).  
In this
case, the common procedure is to use additional SWAP gates to  
transfer the states of two distant qubits to nearest neighbors before
performing the CNOTs. However, in our case,
due to the symmetry of GHZ state, one can go from any state to
the GHZ state in two CNOTs without the need of any quantum SWAP.
Thus even if only nearest-neighbors CNOTs are available
for three qubits on a line, still only 4 CNOTs are enough
to go from any state to any other state.  If one starts from $|000\rangle$ and one allows for relabeling of qubits in the final state,
3 CNOTs are still enough to go to any state,
except for
the GHZ-type class 3 states, Eq.~(\ref{eq:3GHZ}), which need an additional CNOT
in this architecture.

Any two-qubit gate can be expressed in terms of CNOTs and one-qubit
gates.  Thus our result will imply a bound in the number of two-qubit
gates needed to go from one three-qubit state to another,
for any other choice of universal two-qubit gate. We
note that another popular two-qubit gate is the iSWAP
 which is natural for
implementations corresponding to a XY-interaction.
As the iSWAP can be expressed in terms of one CNOT and one SWAP
gate plus one-qubit gates \cite{SchSie03}, our results 
apply directly
to this particular gate 
provided the SWAP can be made classically:
 the number of iSWAPs needed to transform three qubits
is then the same as for CNOTs.
This in particular arises when the physical 
implementation allows for a coupling between any pair of qubits,
as swapping two qubits is equivalent to relabeling the qubits
for all subsequent gates
by interchanging the role of the qubits.
An important example is the
case of superconducting qubits coupled to each other via cavity
bus \cite{MajChoGam07}, one of the most promising recent developments,
where the resonance can be
tuned to couple any pair.  

In conclusion, we have shown that one needs only three CNOTs plus
additional one-qubit gates to transform $|000\rangle$ to any pure three-qubit
states.  If one starts from the GHZ state, only two CNOTs are enough, 
and thus one needs only four CNOTs plus additional one-qubit gates
to transform any initial pure three-qubit state to any
other pure three-qubit states.  An interesting open question is to find out
 whether four is the maximal distance between two three-qubit states
(it should be at least three since $|000\rangle$
is at distance 3 from class 3 states).  It would be also
interesting to know how these results translate to mixed states,
or to pure states using SLOCC instead of LU. 
At last, generalizations to higher numbers
of qubits may pave the way to better optimization of quantum algorithms, 
which are usually described as unitary operators but
 are sometimes just transformations of a given state into another one.

We thank 
the French ANR
(project INFOSYSQQ) and the IST-FET program of the EC
(project EUROSQIP) for funding.
M\v Z would like to acknowledge support by Slovenian Research Agency, grant J1-7437, and hospitality of Laboratoire de Physique Th\' eorique, Toulouse, where this work has been started.


\begin{thebibliography}{99}
\bibitem{relabeling}
In the proofs, qubits are 
relabeled when this simplifies the notations,
so that, e.g., state $\ket{\alpha}_1 \ket{\chi}_{23}$ also represents
$\ket{\chi}_{12}\ket{\alpha}_3$) and 
$\ket{\alpha}_2 \ket{\chi}_{13}$.  As we deal with distances 
from symmetric states such as $\ket{000}$ and GHZ, this amounts to 
relabeling the gates, without any loss of generality.
\bibitem{markoremark}
Indeed, the rank of the reduced density matrix $\rho_{23}$ obtained by tracing out
the first qubit is at most two. 
If it is two, then the range of $\rho_{23}$ contains at least one product vector~\cite{Grassl97}.
If it contains only one, one has class 3(i) states, otherwise one has class 3(ii) 
or class 2 states.
\end{thebibliography}
\end{document}